\documentclass[12pt,a4paper]{article} 

\usepackage[utf8]{inputenc} 

\usepackage{bbm}
\usepackage{amssymb}
\usepackage{graphicx}%
\usepackage{dcolumn}%
\usepackage{bm}%
\usepackage{multirow}
\usepackage{afterpage}
\usepackage{float}
\usepackage[thinlines]{easytable}

\usepackage{geometry} 
\usepackage{setspace}
\usepackage{fullpage}
\usepackage{graphicx} 

\usepackage{enumitem}
\usepackage{amsmath}
\usepackage{bm}
\usepackage{booktabs} 
\usepackage{array} 
\usepackage{paralist} 
\usepackage{verbatim} 
\usepackage{subfig} 
\usepackage{mathtools}

\usepackage{fancyhdr} 
\usepackage{sectsty}
\allsectionsfont{\sffamily\mdseries\upshape} 

\usepackage[nottoc,notlof,notlot]{tocbibind} 
\usepackage[titles,subfigure]{tocloft} 

\usepackage{hyperref}

\begin{document}

\title{
A Thermodynamic Picture of\\ 
Financial Market and Model Risk 
}

\author{Yu Feng\footnote{\emph{Email address}: \href{mailto:fyfly88@gmail.com}{fyfly88@gmail.com}}}


\maketitle

\abstract{By treating the financial market as a thermodynamic system, we establish a one-to-one correspondence between thermodynamic variables and economic quantities. Measured by the expected loss under the worst-case scenario, financial risk caused by model uncertainty is regarded as a result of the interaction between financial market and external information sources. This forms a thermodynamic picture in which a closed system interacts with an external reservoir, reaching its equilibrium at the worst-case scenario. The severity of the worst-case scenario depends on the rate of heat dissipation, caused by information sources reducing the entropy of the system. This thermodynamic picture leads to simple and natural derivation of the characterization rules of the worst-case risk, and gives its Lagrangian and Hamiltonian forms. With its help financial practitioners may evaluate risks utilizing both equilibrium and non-equilibrium thermodynamics.}

\section{Financial Market as a Thermodynamic System}

Financial model risk management deals with the possible losses due to model uncertainty. A non-parametric approach has been proposed to quantify the risk \cite{feng2019non, glasserman2014robust}.  The approach starts with a nominal model that assigns probabilities to different states and then searches for alternative probabilities constrained by an entropic budget. 
The underlying logic of this approach is that the market participants (regarded as a fictitious adversary in \cite{glasserman2014robust}) attempts to change the probabilities using their entropic budget. 

This entropic measure of model risk has a clear thermodynamic interpretation. 
The nominal model can be regarded as a unperturbed thermodynamic system that stays in equilibrium, thus maximizing the entropy. 
The fictitious adversary may be modelled as an external thermal reservoir that interacts with the thermodynamic system. From an information-theoretic point of view, the market participants bring information into the market, creating an information inflow, thus reducing the entropy of the thermodynamic system. As a result, the system may dissipate heat into the external reservoir.

Financial risk managers concern about the worst-case scenario. 
To define the worst-case scenario, we need to model the loss by a loss functional, which is a functional of the path of the relevant risk factors. For instance, the loss of an investment strategy is determined by the path of stock prices within the investment horizon. In the thermodynamic interpretation, each possible path is regarded as a microstate, which has energy that is negatively related to the value of the loss functional. The nominal model assigns probabilities to all microstates, corresponding to a statistical ensemble that represents the unperturbed thermodynamic system. By introducing market participants that perturb the market 
with their own information input, we let the thermodynamic system interact with the external reservoir. In the worst-case scenario, interaction leads to heat dissipation and lower internal energy. This corresponds to a larger expected loss referred as the worst-case risk.

By modeling the financial market as a thermodynamic system, we link the model risk directly with the amount of entropy flowing out of the system. 
At a given entropy, the worst-case scenario corresponds to thermal equilibrium, for in equilibrium the internal energy reaches its minimum (thus expected loss reaches its maximum). This point will be further discussed in Section 3. Assuming information flows into the market continuously, 
we may model the dynamics of the system by a quasi-static process preserving thermal equilibrium. This results in 
a sequence of worst-case scenarios at different levels of entropy.

Overall, to model the worst-case scenario we may regard the financial market as a thermodynamic system that maintains its internal equilibrium. It dissipates heat to external reservoir as a result of information inflow, 
with the reduced internal energy reflecting the worst-case risk. 
The stronger the thermodynamic system interacts with the reservoir, the faster its entropy drops and hence its internal energy. 
From an information-theoretic point of view, the faster new information is brought into the market, the larger the model risk is. Assuming a constant flow of information, a longer time horizon leads to increased reduction in both entropy and internal energy. Therefore higher entropic budget and worst-case risk are expected.


\section{Model Risk and Thermodynamic Variables}

In the theory of dynamic model risk measurement \cite{feng2019non}, we starts from a loss functional $\ell:~\Omega\to\mathbb{R}$ that assigns loss to each state (sample path) $\omega\in\Omega$. We try to find the worst-case scenario due to uncertainty of the market model. Change of the ``true" market model leads to a positive relative entropy between the probabilities before and after the change, reflecting the amount of new information that is brought to the market. Assuming this relative entropy is bounded by a budget $\eta$, we aim to find out the worst-case scenario that maximizes the expected loss, termed as the worst-case risk (primal formulation). 
By introducing a Lagrange multiplier $\theta$, we may solve the problem instead by maximising the entropy penalized risk $W$ (dual formulation). 
The dual formulation results in a value process that corresponds to the penalized risk in the worst-case scenario, a worst-case risk 
$V$, and a relative entropy budget $\eta$. In this section, we will build the one-to-one correspondence between these quantities and thermodynamic variables.

It is noted that the thermodynamic picture of path-dependent model risks roots from the statistical field theory, in which the microstates are expressed through field configurations over $\Omega$. 
As discussed in the previous section, the energy assigned to each microstate is negatively related to the loss functional. We thus assign energies by $\varepsilon(\omega)=-\ell(\omega)$. The macrostate, given by a statistical ensemble of microstates, has an internal energy of
\begin{align}\label{eq:U}
U=
\langle\varepsilon\rangle=-
\langle \ell\rangle
\end{align}
Eq.~\ref{eq:U} links the internal energy $U$ to the expected loss. 
To accommodate the probability measure $P$ associated with the nominal model, we further assign a density of states to each $\omega\in\Omega$. 
Assuming the thermodynamic system consists of $N$ microstates, the density of states is given by
\begin{align}
dn(\omega)=NdP(\omega)
\end{align}

Given the energy and density configurations of all microstates, the internal energy merely depends on the probabilities of states, 
\begin{align}
U=&\int_{\omega\in\Omega}f(\omega)\varepsilon(\omega)dn(\omega)
\end{align}
where $f(\omega)$ is the probability for the system staying in the microstate $\omega\in\Omega$. Normalization condition gives
\begin{align}
\int_{\omega\in\Omega}f(\omega)dn(\omega)=1
\end{align}
In the theory of financial model risk \cite{feng2019non}, $m(\omega)=Nf(\omega)$ is the Radon-Nikodym derivative that transforms the nominal measure $P$ to an equivalent measure $Q(m)$. In the language of thermodynamics, every probability measure represents a statistical ensemble.

The entropy of the thermodynamic system is defined by
\begin{align}
S=&-k\int_{x\in\Omega}f(\omega)\ln f(\omega)dn(\omega)
\end{align}
where $k$ denotes the Boltzmann constant. To link the thermodynamic entropy to the relative entropy $\eta$ in the theory of model risk \cite{glasserman2014robust}, we substitute $f(\omega)$ with $N^{-1}m(\omega)$ and get
\begin{align}
S
=-k\int_{\omega\in\Omega}m(\omega)\ln m(\omega)dP(\omega)+k\ln N
=-k\eta+k\ln N
\end{align}

As discussed in the previous section, the worse-case model may be regarded as the thermodynamic system dissipating heat quasi-statically. Under the dual formulation \cite{feng2019non,glasserman2014robust}, this quasi-statistical process may be characterized by the Lagrange multiplier $\theta$, which takes the role of the inverse temperature $\beta$:
\begin{align}
\theta=\beta:=\frac{1}{kT}
\end{align}
By fixing the temperature $T$, the thermodynamic system undergoes an isothermal process. Furthermore, volume change is not necessary in modeling the financial market. The spontaneity of any isothermal process in a fixed-volume system may be given by the (Helmholtz) free energy 
\begin{align}\label{eq:free}
A=U-TS
\end{align}
In the language of financial model risk, $A$ corresponds to the worst-case risk penalized by the relative entropy, defined by $W:=\langle v\rangle-\theta^{-1}\eta$, in the following way:
\begin{align}
A=-W-\frac{\ln N}{\beta}
\end{align}
Table 1 lists the one-to-one correspondence between thermodynamic variables and quantities in the theory of financial model risk.

\begin{center}\label{tab}
\captionof{table}{Thermodynamic variables and model risk quantities}  
\begin{tabular}{l|l}
\hline\hline\\
Thermodynamic Variable & Quantities in Financial Model Risk\\
\hline\\
Microstate $\omega$ & Sample path $\omega$\\[6pt]
Energy of microstate $\varepsilon(\omega)=-\ell(\omega)$ & Loss functional $\ell(\omega)$\\[6pt]
Density of states $dn(\omega)=NdP(\omega)$ & Nominal probability measure $P(\omega)$\\[6pt]
Thermodynamic ensemble & Probability measure\\[6pt]
Probability of microstate $f(\omega)=N^{-1}m(\omega)$ & Radon-Nikodym derivative $m(\omega)$\\[6pt]
Internal energy $U=-V$ & Expected loss $V=\langle \ell\rangle$\\[6pt]
Entropy $S=-k\eta+k\ln N$ & Relative entropy $\eta$\\[6pt]
Inverse temperature $\beta=(kT)^{-1}$ & Lagrange multiplier $\theta$\\[6pt]
Free energy $A=-W-\beta^{-1}{\ln N}$ & Penalised risk $W=V-\theta^{-1}\eta$\\[6pt]
\hline\hline
\end{tabular}
\end{center}

\section{Worst-case Scenario and Canonical Ensemble}

Different probability measures correspond to different statistical ensembles. 
The main concept here is that the worst-case scenario always corresponds to the ensemble that represents the system in internal thermal equilibrium. 
This conclusion has simple but deep thermodynamic explanations.

Financial model risk is formulated with a primal formulation and its dual formulation \cite{feng2019non, glasserman2014robust}. 
Under the primal formulation the worst-case scenario maximises the expected loss constrained by the entropic budget. The worst-case scenario in this case corresponds to internal equilibrium, as for a given amount of internal energy the maximum entropy is reached in equilibrium. In fact, if the minimum internal energy is given by an non-equilibrium state, it could first transit to the equilibrium state and then dissipate heat to reach a lower level of internal energy. Therefore, the minimum internal energy with entropy bounded from below has to be reached by an equilibrium state.

One can alternatively consider the dual problem in which the penalized model risk $W$ is to be minimized. Given the Lagrange multiplier $\theta$, the worst-case scenario corresponds to the equilibrium ensemble at temperature $T=(k\theta)^{-1}$. 
In the context of thermodynamics, the inverse temperature $\beta$ takes the role of the Lagrange multiplier, converting the constrained problem into a unconstrained minimization of the free energy. At constant temperature and volume any spontaneous process (in an isolated system) tends to reduce the free energy $A$ until reaching equilibrium. Therefore $A$ reaches its minimum at equilibrium, resulting in the maximum of the penalized risk $W$. 


The two formulations of financial model risk correspond to the two forms of the second law of thermodynamics (Table 2). In fact, the primal formulation aims to find the minimum internal energy $U$ of the system at a given entropy $S$. It has a thermodynamic picture of an irreversible isentropic process which ultimately reaches equilibrium. In this picture, we formulate the second law of thermodynamics by $\delta Q< TdS$ where $\delta Q$ is the heat absorbed by the system. In an isentropic process $dS=0$ and therefore $\delta Q< 0$ meaning that the system can only dissipate heat. The internal energy $U$ reaches its minimum at the final state of the isentropic process, and thus the system is in equilibrium.

The dual formulation of financial model risk aims to minimize the free energy $A$ without any constraint. It has a thermodynamic picture of an irreversible isothermal process. At constant temperature we may formulate the second law of thermodynamics by $dA<0$. Therefore, $A$ reaches its minimum at the final state of the isothermal process, and again the system is in equilibrium.
\begin{center}
\captionof{table}{Two formulations of model risk and thermodynamics}  
\begin{tabular}{l|cc}
\hline\hline\\
& Primal problem & Dual problem\\
\hline\\
Formulation &  given $S$, $\min~U$ & $\min~A$\\[8pt]
Thermodynamic process & Irreversible isentropic process & isothermal process\\[8pt]
Second law of thermodynamics & $dU=\delta Q< TdS$ & $dA< 0$\\
 & thus $dS=0\Rightarrow dU< 0$ & \\
\hline\hline
\end{tabular}
\end{center}


\bigskip
In thermal equilibrium the collection of system states is described by the canonical ensemble (or the NVT ensemble). The probability for the system staying in a microstate of energy $\varepsilon$ is given by the Boltzmann distribution
\begin{align}
f_{eq}=\frac{e^{-\beta\varepsilon}}{Z}
\end{align}
where the partition function is defined by
\begin{align}
Z
=\int_{x\in\Omega} e^{-\beta{\varepsilon}}dn(x)=N\mathbb{E}\left( e^{-\beta{\varepsilon}}\right)
\end{align}
The Boltzmann distribution gives the internal energy: 
\begin{align}\label{eq:defueq}
U_{eq}=\frac{1}{Z}\int_{x\in\Omega} \varepsilon e^{-\beta{\varepsilon}}dn(x)
=\frac{N}{Z}\mathbb{E}\left(\varepsilon e^{-\beta{\varepsilon}}\right)
\end{align}
In equilibrium the free energy is a function of the partition function $Z$:
\begin{align}\label{eq:deffree}
A_{eq}=-kT\ln Z
\end{align}
According to the definition of the free energy (Eq.~\ref{eq:free}), the entropy
\begin{align}\label{eq:defseq}
S_{eq}=\frac{U_{eq}-A_{eq}}{T}
=\frac{N}{ZT}\mathbb{E}\left(\varepsilon e^{-\beta{\varepsilon}}\right)
+k\ln Z
\end{align}

Eqs.~\ref{eq:defueq}-\ref{eq:defseq} provide the values of thermodynamic variables when the system is in thermal equilibrium with an external reservoir at temperature $T$. We have shown that this equilibrium corresponds to the worst-case scenario in financial markets, where participants or counter-parties, represented by the external reservoir, bring in new information thus reducing the entropy of the system. Therefore, the values of these thermodynamic variables characterize the worst-case scenario. 
Using the one-to-one correspondence between thermodynamic variables and model risk quantities listed in Table 1, we get the characterizing quantities of the worst-case scenario, including the worst-case risk $V^*$ (defined as the expected loss $V$ under the worst-case scenario) and the relative entropy $\eta^*$ (Table 3).

\begin{center}\label{tab}
\captionof{table}
{Correspondence between equilibrium and worst-case scenario}
\begin{tabular}{l|l}
\hline\hline\\
Thermodynamic Variable & Quantities in Financial Model Risk\\
\hline\\
Thermodynamic equilibrium & Worst-case scenario\\
Probability of microstate 
 \(\displaystyle f_{eq}=\frac{e^{-\beta\varepsilon}}{Z} \)
 & Radon-Nikodym derivative 
 \(\displaystyle m^*=\frac{e^{\theta \ell}}{\mathbb{E}\left( e^{\theta \ell} \right)} \)\\
Internal energy \(\displaystyle U_{eq}
=\frac{N}{Z}\mathbb{E}\left(\varepsilon e^{-\beta{\varepsilon}}\right) \) 
& Worst-case risk \(\displaystyle V^*=\frac{\ell e^{\theta \ell }}{\mathbb{E}\left( e^{\theta \ell } \right)} \)\\[13pt]
Free energy \(\displaystyle A_{eq}=-kT\ln Z\) & Penalised risk \(\displaystyle W^*=\frac{\ln \mathbb{E}\left( e^{\theta \ell } \right)}{\theta}\)\\[11pt]
Entropy  \(\displaystyle S_{eq}
=\frac{N}{ZT}\mathbb{E}\left(\varepsilon e^{-\beta{\varepsilon}}\right)
+k\ln Z\) & Relative entropy \(\displaystyle\eta^*=\theta\frac{\mathbb{E}\left(\ell  e^{\theta \ell } \right)}{\mathbb{E}\left( e^{\theta \ell } \right)}-\ln \mathbb{E}\left(e^{\theta \ell } \right)\)\\
\\
\hline\hline
\end{tabular}
\end{center}

\medskip
The worst-case model has the following thermodynamic picture: a thermodynamic system (initially at $\beta=0$) representing the financial market (initially described by a nominal model) interacts with an external reservoir that represents the collection of market participants. These participants bring in new information, reducing the entropy of the thermodynamic system. In the worst-case scenario, the system dissipates heat and stays in equilibrium.  
During heat dissipation, its internal energy drops as well as the temperature, along with an outgoing entropy flux $\Delta S$ that represents new information flowing into the market. In a dynamic picture, the system may dissipate heat continuously while remaining its internal equilibrium, thus undergoing a quasi-static process. This quasi-static process represents a continuous sequence of worst-case scenarios when an increasing amount of information flows into the market. 

According to the definition of the free energy, Eq.~\ref{eq:free}, the following differential form holds for any thermodynamic process,
\begin{align}
d U-dA=d(TS)
\end{align}
Since we further requires the heat dissipation to be quasi-static, the fundamental thermodynamic relations hold (assuming constant volume): \cite{schmidt2013expansion}
\begin{align}\label{eq:diff}
\begin{aligned}
dU=&\,TdS\\
dA=&\,-SdT
\end{aligned}
\end{align}
In practice, the first equation of Eq.~\ref{eq:diff} allows us to calculate the model risk quantities by integrating along the path of the quasi-static process:
\begin{align}
\Delta S(\beta_f)=&\int_0^{\beta_f}k\beta dU(\beta)~~\text{Thermodynamics}\label{eq:inte0}\\
\Rightarrow~~~~
\eta(\theta_f)=&\int_0^{\theta_f}\theta dV(\theta)~~~~~\text{Financial model risk}\label{eq:inte}
\end{align}
Eq.~\ref{eq:inte} shows that we may calculate the relative entropy $\eta$ by simulating the quasi-static process. 
Given the worst-case risk $V$ as a function of the Lagrange multiplier $\theta$, we may integrate $\theta$ from 0 to a final value $\theta_f$ to obtain the relative entropy $\eta$ that corresponds to $\theta_f$. 

As a simple example, consider the case of $n$-dimensional ideal gas, in which the density of states is given by
\begin{align}
dn(\varepsilon)=|\varepsilon|^{\frac{n}{2}-1}d\varepsilon
\end{align}
where state energy $\varepsilon\in(-\infty, 0)$ (so that the loss per state $\ell=-\varepsilon$ is positive). Using the equipartition theory, the internal energy as a function of the inverse temperature is  given by
\begin{align}
U(\beta)
=-\frac{n}{2\beta}
\end{align}
According to Eq.~\ref{eq:inte0}, we have
\begin{align}
\Delta S(\beta)=&-\frac{kn}{2}\ln \frac{U(\beta)}{U(0)}~~\,\text{Thermodynamics}\\
\Rightarrow~~~~
\eta(\theta)=&\,\frac{n}{2}\ln\frac{V(\theta)}{V(0)}~~~~~~~~\text{Financial model risk}\label{eq:ig1}
\end{align}
Eq.~\ref{eq:ig1} allows us to express the worst-case risk as a function of the relative entropy for this special example:
\begin{align}\label{eq:veta}
V(\eta)=V(0)\exp\left(\frac{2\eta}{n}\right)
\end{align}
In this example, the worst-case risk, or more accurately the expected loss under the worst-case scenario, increases exponentially with the relative entropy budget $\eta$. The dimension $n$ determines the rate of the exponential growth.

\section{Information Flow, Non-equilibrium Dynamics and Simulated Thermalization}

Under the thermodynamic interpretation of model risk, the nominal model corresponds to an isolated system, representing a financial market that does not incorporate new information. In this ideal case, the market behaves exactly as the nominal model describes, and no model uncertainty exists due to lack of new information. In reality, however, there is no perfect model. Any probabilistic model of the financial market has some uncertainty, and requires frequent recalibration according to new information. The real financial market is more like an open system which interacts with the environment. When new information flows into the system, the entropy of the system is reduced.

To measure the impact of information input, we consider the probability law of the microstates. We represent microstates by a random variable $X$. The thermodynamic entropy $S$ is related to the information entropy $H(X)$ by $S=kH(X)$. Now assuming there are $n$ random variables $Y_i,~i=1,2,\cdots,n$ whose results are material to the market. In financial markets these random variables could represent market events such as financial report releases. 
Incorporating the new information about $Y_i$, we may update the information entropy $H(X)$ by \cite{cover2012elements}:
\begin{align*}
H(X|Y_1,\cdots,Y_n)=H(X)-I(X;Y_1)-I(X;Y_2|Y_1)-\cdots-I(X;Y_n|Y_1,\cdots,Y_{n-1})
\end{align*}
where $I(X;Y_i)$ denotes the mutual information between the two random variables, $X$ and $Y_i$, and $I(X;Y_i|Y_1,Y_2,\cdots,Y_{i-1})$ is the conditional mutual information. Both quantities are non-negative, resulting in a reduction of the information entropy. Statistically, we may express the sequence of $X_i$ by an incoming rate of mutual information $I(t)$. 
The rate of mutual information multiplying the time horizon provides the amount of entropy that can be taken away by new information:
\begin{align}
\Delta S=k\Delta H(X)=-k\int_0^T I(t)dt
\end{align}
Therefore, 
the faster new information is brought into the market, the larger the model risk is. Since financial market always has inflow of information, a longer time horizon leads to a larger entropy reduction. Therefore larger relative entropy budget and larger model risk are expected.

Let us take a look back at the quasi-static process, which models a sequence of worst-case scenarios when information flows into the market continuously. The integrated form Eq.~\ref{eq:inte} describes such quasi-static heat dissipation. Given $V(\theta)$, it provides a convenient way of obtaining the relation between the worst-case risk $V$ and the relative entropy $\eta$. However, in general a closed-form expression for $V(\theta)$ may not be available. 
We need to find an efficient numerical routine of calculating the function $V(\theta)$ in such cases. 
This can be done by simulating the non-equilibrium transition from the unperturbed state (the nominal model) to the new equilibrium state (the worst-case model or scenario) after an instantaneous extraction of internal energy. 

We may simulate the non-equilibrium transition using an algorithm called simulated thermalization. This algorithm can be categorized into the ensemble Monte Carlo methods. 
Rather than calculating the worst-case risk $V$ for a given $\theta$, it calculates $\theta$ for a given $V$. In practice, it allows financial practitioners to find the value of the Lagrangian multiplier required to reach a given level of risk. 
The concept of simulated thermalization is to simulate the dynamics of ensemble 
by considering merge and creation of two microstates at different energy levels. 
Following are the steps of the algorithm:
\begin{description}
\item[(1)] initialize $N$ evenly distributed energy levels $\varepsilon_i$ ($i=1,2,\cdots,N$) with their respective density of states $n_i$ according to the nominal probability measure
\item[(2)] for a given $V$, initialize the state probabilities $f_i$ such that the total energy equals $-V$
\item[(3)] randomly pick three energy levels $\varepsilon_i$, $\varepsilon_j$ and $\varepsilon_k$ such that $\varepsilon_i=\varepsilon_j+\varepsilon_k$, followed by calculating the flow proportional to $(f_i-f_jf_k)n_in_jn_k$ (i.e. net transition rate for a particle at $\varepsilon_i$ to decay into two particles with energies $\varepsilon_j$ and $\varepsilon_k$)
\item[(4)] adjust the three probabilities $f_i$, $f_j$ and $f_k$ according to the flow / transition. Remember to set a proper adjustment rate (or learning rate) to avoid getting into negative probabilities
\item[(5)] repeat steps 3 and 4 until the average particle energy, given by $-V/\sum_{i=1}^Nf_in_i$, converges. The probabilities $f_i$ should now follow an exponential law, i.e. $f_i\sim e^{-\beta\varepsilon_i}$, where the inverse temperature $\beta$, and hence the Lagrange multiplier $\theta$, can be immediately obtained.
\end{description}

The simulated thermalization algorithm allows the initialized ensemble to gradually transit to equilibrium. The convergence rate of this algorithm follows the non-equilibrium dynamics of statistical ensemble. 
In non-equilibrium statistical mechanics, the transition rates among three ensembles I, II and III follow the relation as below (assuming no external work): \cite{perunov2016statistical}
\begin{align}\label{eq:25}
\ln\left(\frac{I\to II}{II\to I}\right)-
\ln\left(\frac{I\to III}{III\to I}\right)
\approx
\left\langle\ln\frac{p}{p_{bz}}\right\rangle_{III}-
\left\langle\ln\frac{p}{p_{bz}}\right\rangle_{II}+
\beta\ln\frac{Z_{II}}{Z_{III}}
\end{align}
where $X\to Y$ is the net transition rate between two statistical ensembles, $X$ and $Y$. $\beta$ is the inverse temperature of the external reservoir that interacts with the system, and $p_{bz}$ denotes the Boltzmann distribution of microstates corresponding to the canonical ensemble. 
In the special case where ensemble III refers to thermal equilibrium (with the external reservoir), Eq.~\ref{eq:25} is simplified to
\begin{align}\label{eq:26}
\ln\left(\frac{I\to II}{II\to I}\right)-
\ln\left(\frac{I\to eq}{eq\to I}\right)
\approx
-\left\langle\ln\frac{p}{p_{bz}}\right\rangle_{II}
+\beta\ln\frac{Z_{II}}{Z_{eq}}
\end{align}
We simplify Eq.~\ref{eq:26} by further assuming that the two ensembles I and II refer to the same non-equilibrium ensemble. This results in an expression for the net (logarithmic) transition rate between a non-equilibrium ensemble and the canonical ensemble:
\begin{align}\label{eq:neq}
\ln\left(\frac{neq\to eq}{eq\to neq}\right)
\approx&
\left\langle\ln\frac{p}{p_{bz}}\right\rangle_{neq}
-\beta\ln\frac{Z_{neq}}{Z_{eq}}\\
=&\,D(p\,||\,p_{bz})-\beta\ln\frac{Z_{neq}}{Z_{eq}}
\end{align}
where $D(p\,||\,p_{bz})$ denotes the relative entropy between the non-equilibrium distribution $p$ and the Boltzmann distribution $p_{bz}$. It is always non-negative. When the system is in thermal equilibrium, the two distributions are identical resulting in a zero relative entropy, and the net transition rate only depends on the second term on the RHS of Eq.~\ref{eq:neq}. 
According to  Eq.~\ref{eq:deffree}, when the system is in internal equilibrium this term is the difference between free energies. The spontaneous transition always points to the state with lower free energy. 
Therefore, the transition, with rate given by Eq.~\ref{eq:neq}, converges when the system reaches internal equilibrium, and its free energy decreases to minimum by interacting with the external reservoir. Characterized by the transition rate, Eq.~\ref{eq:neq}, the convergence rate of the simulated thermalization algorithm is approximately linear.

\section{Lagrangian and Hamiltonian formulations}

It is well known that statistical field theory is closely related to quantum field theory via Wick rotation \cite{itzykson1991statistical}.
Quantum field theory can be formulated using either Hamiltonian operator or equivalently Feynman's path integral. 
Following the work on formulating derivative pricing problems using Hamiltonian operator and path integral \cite{baaquie2007quantum}, we apply the same approach to the problem of financial model risk. 
In a similar way to quantum field theory, the pricing kernel (or conditional probability) can be given by both Hamiltonian operator and path integral 
\begin{align}
p(x,T;x_0,0)=\frac{\langle x,T|e^{-\int_0^T\hat{H}_0d\tau}|x_0,0\rangle}{Z_0}=\frac{1}{Z_0}\int\mathcal{D}X e^{S_0}
\end{align}
where $\hat{H}_0$ and $S_0$ are the unperturbed Hamiltonian and the unperturbed action, respectively, both characterizing the conditional probabilities in the nominal model. 
It is noted that a state is described by $|x,\tau\rangle$ where $\tau=T-t$ is the time left in an investment horizon. 
In a nominal model that relies on a driftless process with constant diffusion $\sigma$, we have \cite{baaquie2007quantum}
\begin{align}
\hat{H}_0=-\frac{\sigma^2}{2}\frac{\partial^2}{\partial x^2}~~~~\text{and}~~~~S_0=-\int \frac{1}{2\sigma^2}\left(\frac{dx}{d\tau}\right)^2d\tau
\end{align}
The partition function is given by
\begin{align}
Z_0=&\sum_x\langle x,T|e^{-\int_0^T\hat{H}_0d\tau}|x_0,0\rangle
\end{align}
It is noted that the name ``pricing kernel'' originates from the problem of derivative pricing. \cite{baaquie2007quantum} Since it is merely a characterization of the underlying stochastic process, 
we may use the same terminology in the context of financial model risk. 

The worst-case scenario corresponds to the canonical ensemble at a given temperature. Physically it describes a system in equilibrium with an external reservoir at the same temperature. The total Hamiltonian operator in this case is
\begin{align}\label{eq:tH}
\hat{H}=\hat{H}_0+\beta\hat{H}_I
\end{align}
where 
the interaction Hamiltonian $\hat{H}_I$ describes 
the interaction with the thermal bath, and affects the total Hamiltonian via the inverse temperature $\beta$. 
In thermal quantum field theory, it represents a quantum system in equilibrium with a thermal bath at the inverse temperature $\beta$ \cite{khanna2009thermal}. 
The path integral formation of the pricing kernel is therefore given by a total action $S$:
\begin{align}\label{eq:33}
q(x,T;x_0,0)=
\frac{1}{Z}\int\mathcal{D}X e^{S}
\end{align}
where $S$ is the sum of the unperturbed action and an interaction action
\begin{align}\label{eq:34}
S=S_0+ \beta S_I
\end{align}
The interaction action is related to the Hamiltonian operator by
\begin{align}\label{eq:action}
e^{S_0+\beta S_I}=\lim_{N\to\infty}\Pi_{i=1}^N\langle x_i,\tau_i|e^{-(\tau_i-\tau_{i-1})\hat{H}}|x_{i-1},\tau_{i-1}\rangle
\end{align}

According to the definition of the Hamiltonian operator \cite{baaquie2007quantum}, we have
\begin{align}\label{eq:HV}
\frac{\partial C}{\partial t}=-\frac{\partial C}{\partial\tau}=\hat{H}C
\end{align}
where $C=\mathbb{E}_t(g(x_T))$ is the conditional expectation of some terminal function (i.e. function of the terminal state $x_T$), $g(x_T)$, under the probability measure characterized by the Hamiltonian $\hat{H}$. When the Hamiltonian takes the form of Eq.~\ref{eq:tH}, it characterizes the worst-case scenario.
This inspires us to derive the partial differential equation that governs the financial model risk \cite{feng2019non} using a quantum mechanical approach. We may first express the loss functional using the Lagrangian formulation and then converts to its corresponding Hamiltonian formulation. In fact, the interaction action $S_I$ is given by the loss functional $\ell$, for according to Eqs.~\ref{eq:33} and \ref{eq:34} the probability of a path is proportional to $e^{\beta S_I}$. According to the expression for $m^*$ in Table 2, it is clear that $S_I$ should take the value of the path-wise loss $\ell$. 

Now consider the following loss functional as an example:
\begin{align}\label{eq:specialvec}
\ell= \int_0^T{h}(t)dx_t+g(x_T)
\end{align}
We may define the interaction Lagrangian by
\begin{align}\label{eq:L}
\mathcal{L}_I={h}\frac{dx}{dt}=-{h}\frac{dx}{d\tau}
\end{align}
We can show that its corresponding Hamiltonian formulation takes the form of
\begin{align}\label{eq:H}
\hat{H}=
\hat{H}_0-\beta\sigma^2{h}\frac{\partial}{\partial {x}}
\end{align}
In fact, we may evaluate a single component of the product Eq.~\ref{eq:action} where $\varepsilon\to 0^+$:
\begin{align}
\langle x,\tau|e^{-\varepsilon \hat{H}}|y,\tau-\varepsilon\rangle&=\int_{-\infty}^\infty e^{-\varepsilon \hat{H}}e^{ip(x-y)}dp\nonumber\\
&= \int_{-\infty}^\infty e^{-\varepsilon(\sigma^2p^2-i\beta\sigma^2hp)}
e^{ip(x-y)}dp\nonumber\\
&=\frac{1}{\sqrt{2\pi\varepsilon\sigma^2}}\exp\left[-\frac{\varepsilon}{2\sigma^2}\left(\frac{x-y}{\varepsilon}+\beta\sigma^2{h}\right)^2\right]\nonumber\\
&\xrightarrow{\varepsilon\to 0^+}\frac{1}{\sqrt{2\pi\varepsilon\sigma^2}}\exp\left(-\frac{\varepsilon \beta^2h^2}{2\sigma^2}\right)
\exp\left[-\frac{\varepsilon}{2\sigma^2}\left(\frac{dx}{d\tau}\right)^2-\varepsilon\frac{dx}{d\tau}\beta{h}\right]\nonumber\\
&=\frac{1}{\sqrt{2\pi\varepsilon\sigma^2}}\exp\left(\frac{\varepsilon \beta^2h^2}{2\sigma^2}\right)
e^{\varepsilon\mathcal{ L}}
\end{align}
where the total Lagrangian
\begin{align}
\mathcal{ L} =\, -\frac{1}{2\sigma^2}\left(\frac{dx}{d\tau}\right)^2-\beta h\frac{dx}{d\tau}
=\,\mathcal{ L}_{0}+\beta\mathcal{L}_I
\end{align}
is the sum of the unperturbed Lagrangian and the interaction Lagrangian. 
This verifies that Eq.~\ref{eq:H} gives the Hamiltonian which corresponds to the interaction Lagrangian in Eq.~\ref{eq:L}.
Substituting Eq.~\ref{eq:H} into Eq.~\ref{eq:HV} we get
\begin{align}
\frac{\partial C}{\partial t}=-\left(\frac{\sigma^2}{2}\frac{\partial^2}{\partial{x}^2}+\beta\sigma^2{h}\frac{\partial}{\partial {x}}\right)C
\end{align} 
for $C=\mathbb{E}_t(g(x_T))$. Now we consider another conditional expectation defined by
\begin{align}
D=\mathbb{E}_t\left(\int_t^Th\frac{dx}{dt}dt\right)
\end{align}
we have $
\frac{\partial D}{\partial x}=0$ and
\begin{align}
\frac{\partial D}{\partial t}=-h\mathbb{E}_t\left(\frac{dx}{dt}\right)
=\lim_{\varepsilon\to 0^+}h\int\langle x,\tau|\frac{x-y}{\varepsilon}e^{-\varepsilon \hat{H}}|y,\tau-\varepsilon\rangle dx
=-\beta\sigma^2 h^2
\end{align}
For the time-dependent worst-case risk defined by $V=C+D$, we get the following partial differential equation:
\begin{align}
\frac{\partial V}{\partial t}=-\left(\frac{\sigma^2}{2}\frac{\partial^2}{\partial{x}^2}+\beta\sigma^2{h}\frac{\partial}{\partial {x}}\right)V-\beta\sigma h^2
\end{align}
or 
\begin{align}
\frac{\partial V}{\partial t}+\beta\sigma^2{h}\left(\frac{\partial V}{\partial {x}}+h\right)+\frac{\sigma^2}{2}\frac{\partial^2V}{\partial{x}^2}=0
\end{align}
subject to $V=g$ when $t=T$ (terminal condition). If replacing the inverse temperature $\beta$ by the Lagrange multiplier $\theta$, we get the partial differential equation that governs the worst-case risk. 
This path-dependent partial differential equation were initially derived using a much more complex approach of functional Ito calculus \cite{feng2019non}.

\bibliographystyle{unsrt}
\bibliography{ref}

\begin{thebibliography}{1}

\bibitem{feng2019non}
Yu~Feng.
\newblock Non-parametric robust model risk measurement with path-dependent loss
  functions.
\newblock {\em arXiv preprint arXiv:1903.00590}, 2019.

\bibitem{glasserman2014robust}
Paul Glasserman and Xingbo Xu.
\newblock Robust risk measurement and model risk.
\newblock {\em Quantitative Finance}, 14(1):29--58, 2014.

\bibitem{schmidt2013expansion}
Klaus Schmidt-Rohr.
\newblock Expansion work without the external pressure and thermodynamics in
  terms of quasistatic irreversible processes.
\newblock {\em Journal of Chemical Education}, 91(3):402--409, 2013.

\bibitem{cover2012elements}
Thomas~M Cover and Joy~A Thomas.
\newblock {\em Elements of information theory}.
\newblock John Wiley \& Sons, 2012.

\bibitem{perunov2016statistical}
Nikolay Perunov, Robert~A Marsland, and Jeremy~L England.
\newblock Statistical physics of adaptation.
\newblock {\em Physical Review X}, 6(2):021036, 2016.

\bibitem{itzykson1991statistical}
Claude Itzykson and Jean-Michel Drouffe.
\newblock {\em Statistical field theory}.
\newblock Cambridge University Press, 1991.

\bibitem{baaquie2007quantum}
Belal~E Baaquie.
\newblock {\em Quantum finance: Path integrals and Hamiltonians for options and
  interest rates}.
\newblock Cambridge University Press, 2007.

\bibitem{khanna2009thermal}
Faqir~C Khanna.
\newblock {\em Thermal quantum field theory: algebraic aspects and
  applications}.
\newblock World Scientific, 2009.

\end{thebibliography}

\end{document}